\documentclass[twocolumn, iop, apj, twocolappendix, numberedappendix, appendixfloats]{emulateapj}
\usepackage{CJKutf8}
\usepackage{threeparttable}
\usepackage{multirow}
\usepackage{times}
\usepackage{enumitem}
\usepackage{url} 
\usepackage{color}
\usepackage{amsmath,bm}
\usepackage{subfigure}
\usepackage{natbib}
\usepackage{graphicx}
\usepackage{graphics}
\usepackage{hyperref}                             
\usepackage{amsmath}
\usepackage{bm}
\usepackage{subfigure}
\usepackage{array}
\usepackage{booktabs}
\usepackage{lineno}

\shorttitle{Galactic Disk North-south Asymmetry in Metallicity}
\shortauthors{Sun et al.}
\submitted{Submitted to ApJL;\,\,\,\,Accepted December 17, 2024}

\begin{document}

\title{The Galactic Disk North-south Asymmetry in Metallicity May Be A New Tracer for the Disk Warp}

\author{Weixiang Sun\textsuperscript{1,5}}
\author{Han Shen\textsuperscript{3,4}}
\author{Biwei Jiang\textsuperscript{1,5}}
\author{Xiaowei Liu\textsuperscript{2,5}}

\altaffiltext{1}{School of Physics and Astronomy, Beijing Normal University, Beijing 100875, People’s Republic of China; {\it sunweixiang@bnu.edu.cn {\rm (WXS)}}; {\it bjiang@bnu.edu.cn {\rm (BWJ)}}}
\altaffiltext{2}{South-Western Institute for Astronomy Research, Yunnan University, Kunming 650500, People's Republic of China; {\it x.liu@ynu.edu.cn {\rm (XWL)}}}
\altaffiltext{3}{School of Physics, University of New South Wales, Kensington 2032, Australia}
\altaffiltext{4}{ARC Centre of Excellence for All Sky Astrophysics in 3 Dimensions (ASTRO 3D), Australia}
\altaffiltext{5}{Corresponding authors}

\begin{abstract}

Galactic disk warp has been widely characterized by stellar distributions and stellar kinematics but has not been traced by stellar chemistry.
Here, we use a sample with over 170,000 red clump (RC) stars selected from LAMOST and APOGEE first to establish a correlation between the north-south asymmetry in metallicity ([Fe/H]) and the disk warp.
Our results indicate that the height of the [Fe/H] mid-plane for the whole RC sample stars is accurately described as $Z_{w}$ = 0.017\,($R$\,$-$\,7.112)$^{2}$\,sin($\phi$\,$-$\,9.218).
This morphology aligns closely with the warp traced by Cepheids, suggesting that the disk north-south asymmetry in [Fe/H] may serve as a new tracer for the Galactic warp.
Our detailed analysis of the young/thin disk stars of this RC sample suggests that its warp is well-modeled as $Z_{w}$ = 0.016\,($R$\,$-$\,6.507)$^{2}$\,sin($\phi$\,$-$\,4.240), indicating that the line of node (LON) of the Galactic warp is oriented at 4.240\,$_{-1.747}^{+1.641}$ degree.

\end{abstract}


\section{Introduction}

The Milky Way is a warped galaxy has been inferred first by the radio observations of neutral gas \citep[e.g.,][]{Kerr1957}, and later confirmed by molecular clouds \citep[e.g.,][]{Grabelsky1987}, interstellar dust \citep[e.g.,][]{Freudenreich1994, Drimmel2001}, and stars \citep[e.g.,][]{Drimmel2001, Lopez-Corredoira2002b, Chen2019}.
The results indicate that the Galactic disk is flat out to the Solar radius, then bends downwards in the south and upwards in the north, with the line-of-node close to the Sun \citep[e.g.,][]{Freudenreich1994, Drimmel2001, Lopez-Corredoira2002b, Momany2006}.

Theoretically, the Galactic warp is typically considered to be shaped by the disk perturbations that may be from (i) intergalactic magnetic fields \citep[e.g.,][]{Battaner1990};
(ii) a misaligned dark halo \citep[e.g.,][]{Sparke1988, Ostriker1989, Debattista1999}; 
(iii) interactions with satellites galaxies \citep[e.g.,][]{Kim2014};
(iv) the cosmic infall \citep[e.g.,][]{Jiang1999};
(v) accretion of intergalactic matter \citep[e.g.,][]{Kahn1959, Lopez-Corredoira2002a}, amongst others.
Currently, the origin of the Galactic warp is still under debate, although several attempts have been made \citep[e.g.,][]{Miyamoto1988, Lopez-Corredoira2014, Schonrich2018, Chen2019}.
One of the keys to resolving this debate is to make a reliable tracking and measuring of the Galactic warp.
However, the direct trace and measure of the Galactic warp are currently difficult achievement, mainly because of the incomplete sampling and selection effect of the samples for the direct measurement of 3D structures of the disk warp (e.g., Momany et al. 2006; Chen et al. 2019), owing to which researchers tend to trace and measure the warp based on the behaviors of the disk stars that are strongly affected by the warp \citep[e.g.,][]{Poggio2018, Huang2018}.

\begin{figure*}[t]
\centering
\subfigure{
\includegraphics[width=8.4cm]{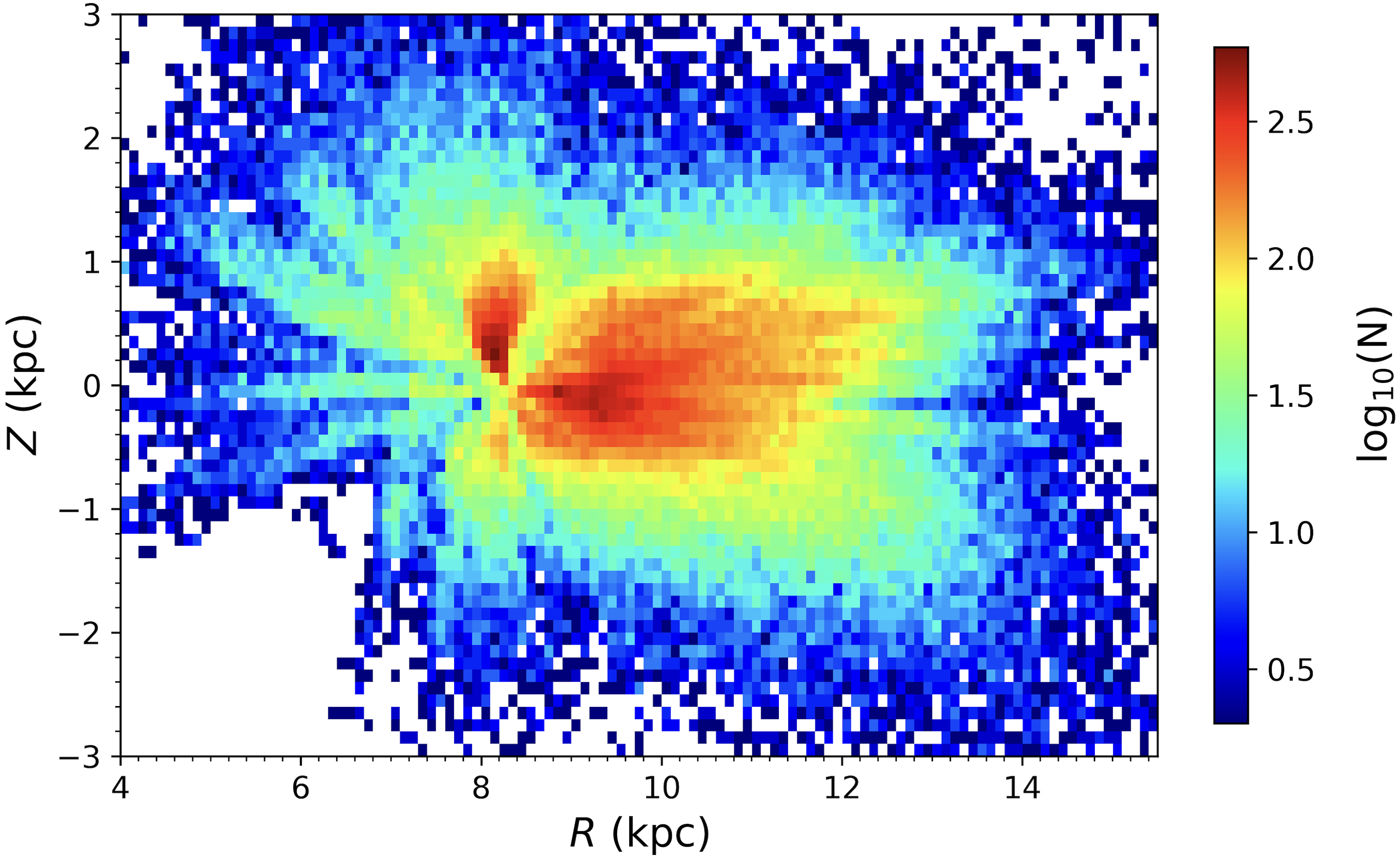}
}
\hspace{0.15cm}
\subfigure{
\includegraphics[width=8.4cm]{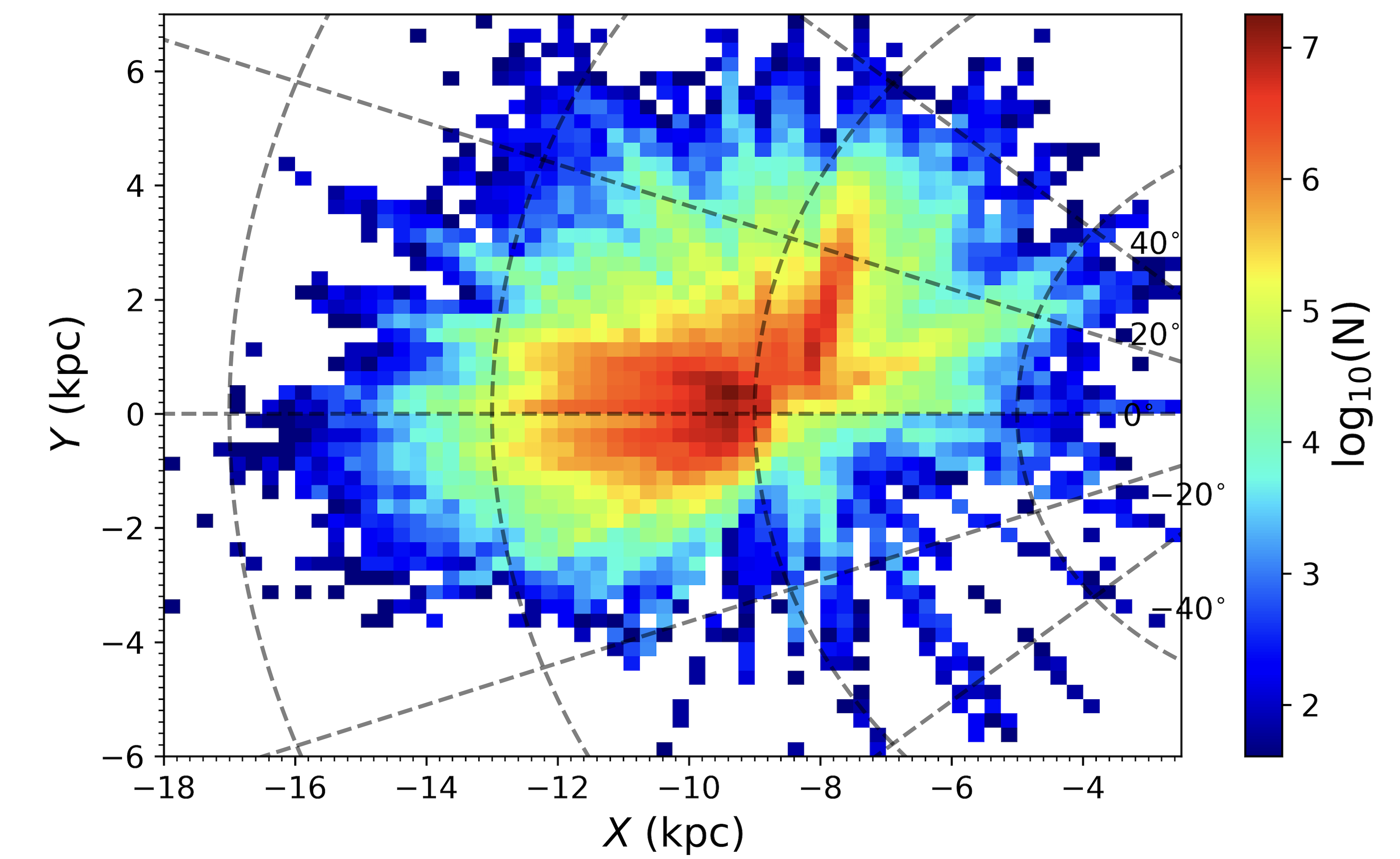}
}

\caption{{\bf Left panel:} Spatial distribution of the LAMOST RC sample stars, in the $R$--$Z$ plane, color-coded by stellar number densities.
There are no less than two stars in a bin, with both axes spaced by 0.1 kpc.
{\bf Right panel:} Spatial distribution of the LAMOST RC sample stars, in the $X$--$Y$ plane, with over-plotted by the $\phi$ angle line.
There is a minimum of five stars per bin, with both axes spaced by 0.25 kpc.
The RC sample stars distributed between 4.0\,$\leq$\,$R$\,$\leq$\,15.0\,kpc and $|Z|$\,$\leq$\,3.0\,kpc, and covered a range from $\phi$ = $-$20 degree to $\phi$ = 40 degree.}
\end{figure*}

\begin{figure}[t]
\begin{center}
\includegraphics[width=8.5cm]{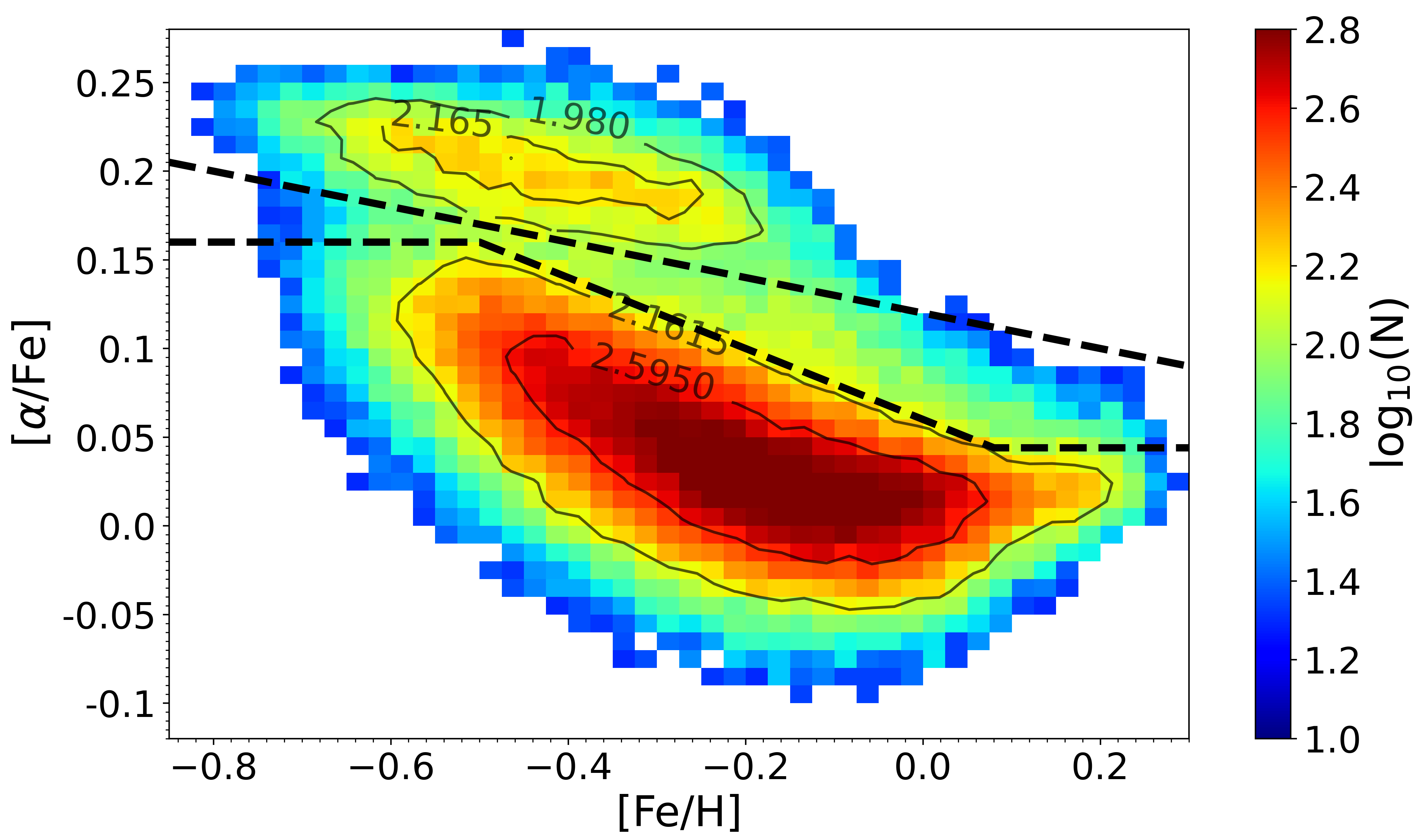}
\caption{Distribution of the RC sample stars, in the [Fe/H]$-$[$\alpha$/Fe] plane, over-plotted with contours of equal densities.
The number densities are represented by the colorbar on the right.
The horizontal axis and vertical axis are respectively spaced by 0.025\,dex and 0.02\,dex, with no less than 20 stars in a bin.
The two dashed lines separate the thin (below the lines) and the thick (above the lines) disk stars.}
\end{center}
\end{figure}

Several studies \citep[e.g.,][]{Gaia Collaboration2018, Huang2018, Li2020, Sun2024a} presented an indirect method to trace and characterize the warp properties based on the stellar kinematics.
Those results provide a deep understanding of the properties of the Galactic warp, such as, the Galactic warp leads the stellar vertical velocity ($V_{Z}$) changes as Galactic radius ($R$) and azimuthal angle ($\phi$), with the $V_{Z}$ increases as $R$ increases out the solar circle \citep[e.g.,][]{Gaia Collaboration2018, Huang2018, Li2020}; and the warp feature may be a long-lived feature in the Milky Way disk \citep[e.g.,][]{Poggio2018, Sun2024a}.
However, tracing the disk warp based on an indirect method with stellar kinematics will lead to a more complex scientific problem, one of which is how to assume a reliable kinematic model.
Indeed, the connection between the structural properties of the warp and its kinematic signature requires a model of how the warp evolves with time.
One cannot derive the 3D shape of the warp from the kinematics only, unless one assumes a specific kinematic model.
Furthermore, the warp parameters (such as the amplitude, onset radius, and LON) are still under hot debate \citep[e.g.,][]{Drimmel2001, Yusifov2004, Amores2017, Chen2019, Li2020}, which are primarily caused by the difficult process of de-projecting the three-dimensional structure of the Galactic disk (due to interstellar extinction and the selection function of the survey).
Therefore, to make a reliable constraining of the parameters of the Galactic warp, it can be useful to establish a new direct tracer for characterizing the nature of the Galactic warp based on its well-known structure.
A recent study has provided inspiration for constructing chemical tracers for the Galactic warp, Sun et al. ({\color{blue}{2024b}}) conducted a detailed measurement of the metallicity distributions of the Galactic thin and thick disks, and found that the thin disk displays an obvious north-south asymmetry in metallicity $R$ larger than around 8$\sim$9.0\,kpc, which is roughly consistent with the disk warp structure in morphology, and suggests that it may be related to the disk warp.
In this paper, we attempt to relate the north-south asymmetry in [Fe/H] of the Galactic disk to the Galactic warp, to establish the first detection of the chemical signature for the Galactic warp, thereby constraining its fundamental parameters.

This paper is structured as follows.
In Section\,2, we describe the data used in this paper,
and present our results and discussion in Section\,3. 
Finally, our main conclusions are summarized in Section\,4.

\section{Data}

\begin{figure*}[t]
\centering
\subfigure{
\includegraphics[width=8.4cm]{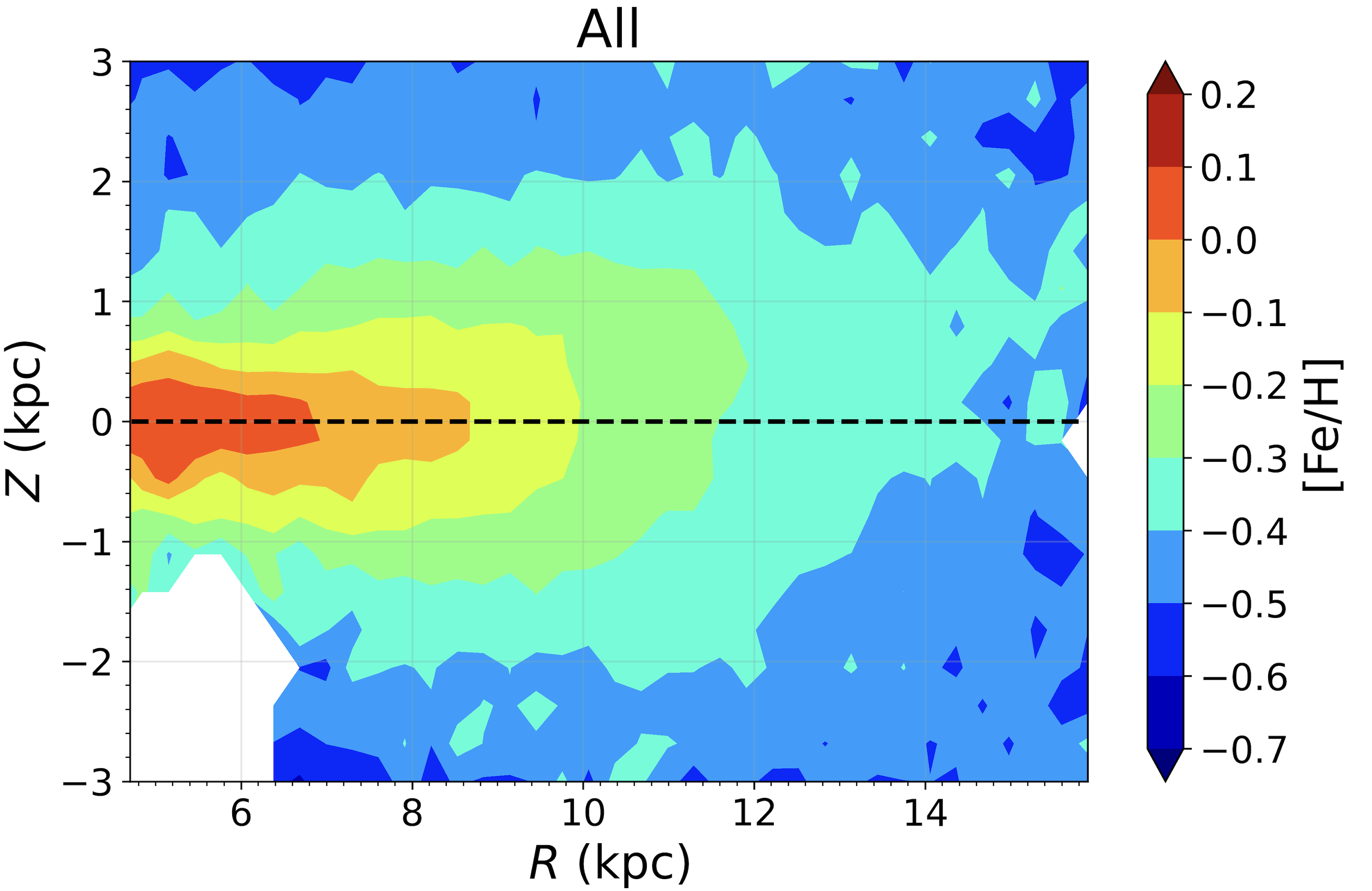}
}
\hspace{0.15cm}
\subfigure{
\includegraphics[width=8.4cm]{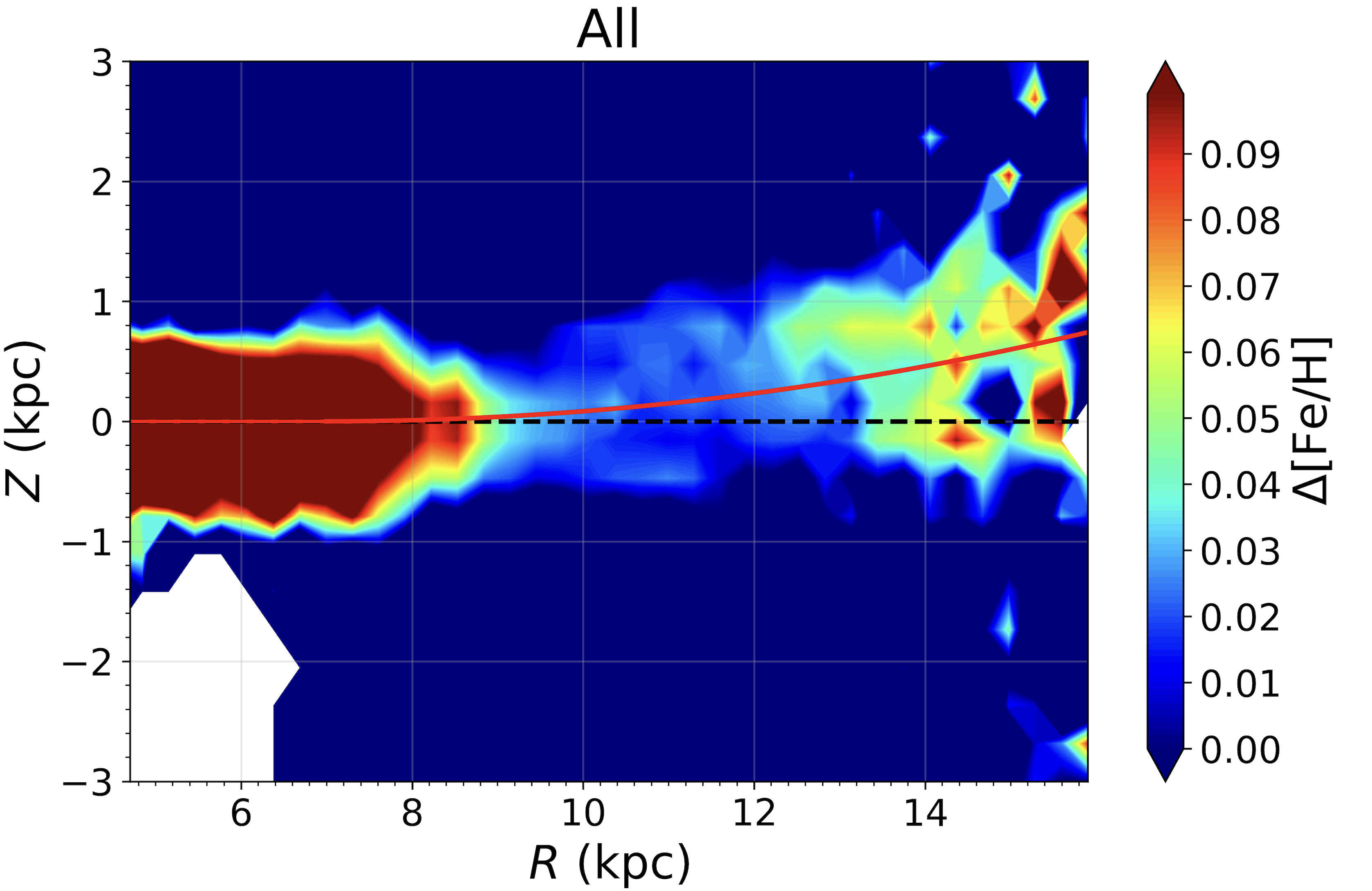}
}

\caption{Metallicity ([Fe/H], left panel) and remaining metallicity ($\Delta$[Fe/H], right panel) distribution, in the $R$ - $Z$ plane, of the whole RC sample stars.}
\end{figure*}

\begin{figure*}[t]
\centering
\subfigure{
\includegraphics[width=8.4cm]{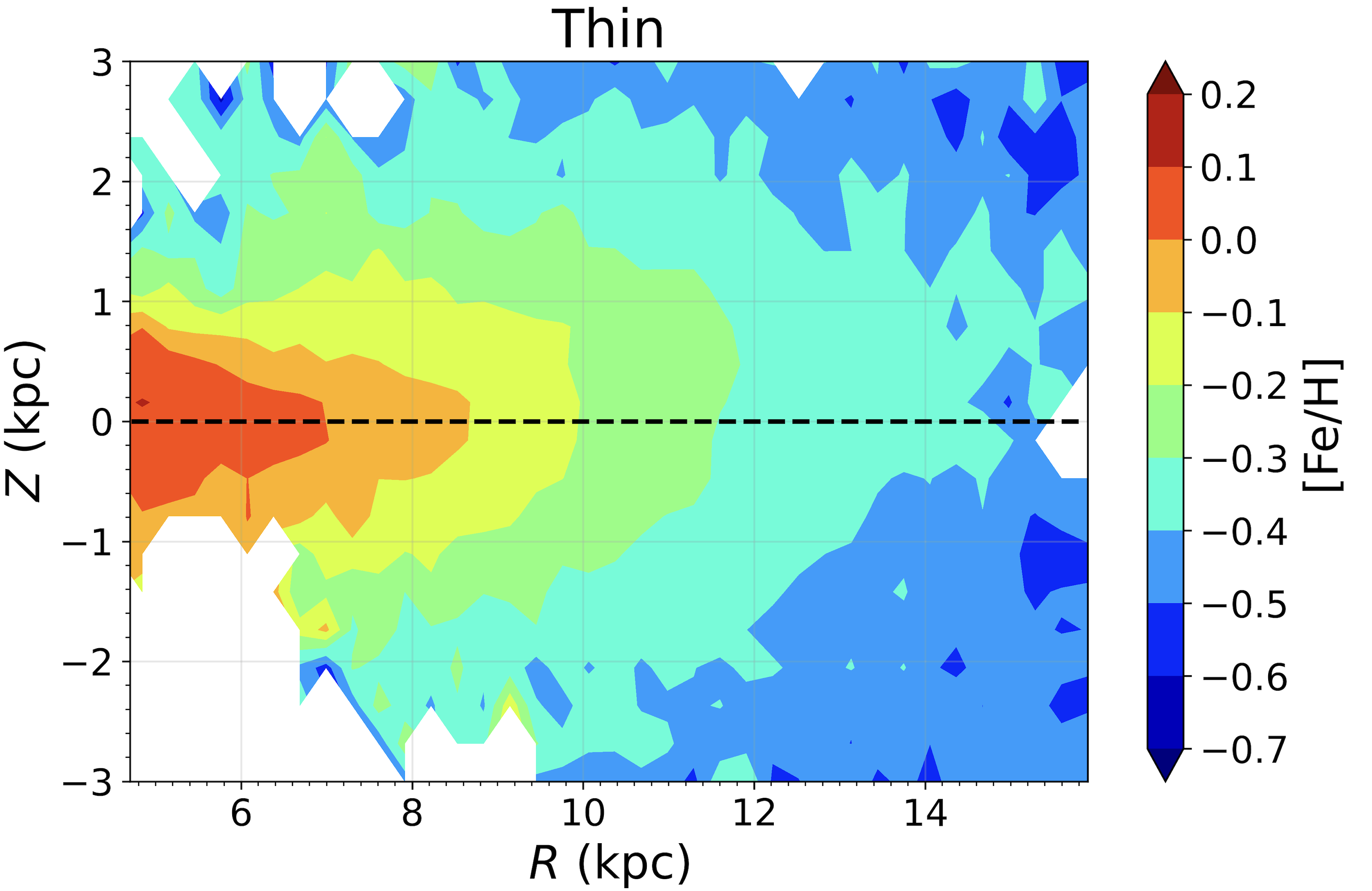}
}
\hspace{0.15cm}
\subfigure{
\includegraphics[width=8.4cm]{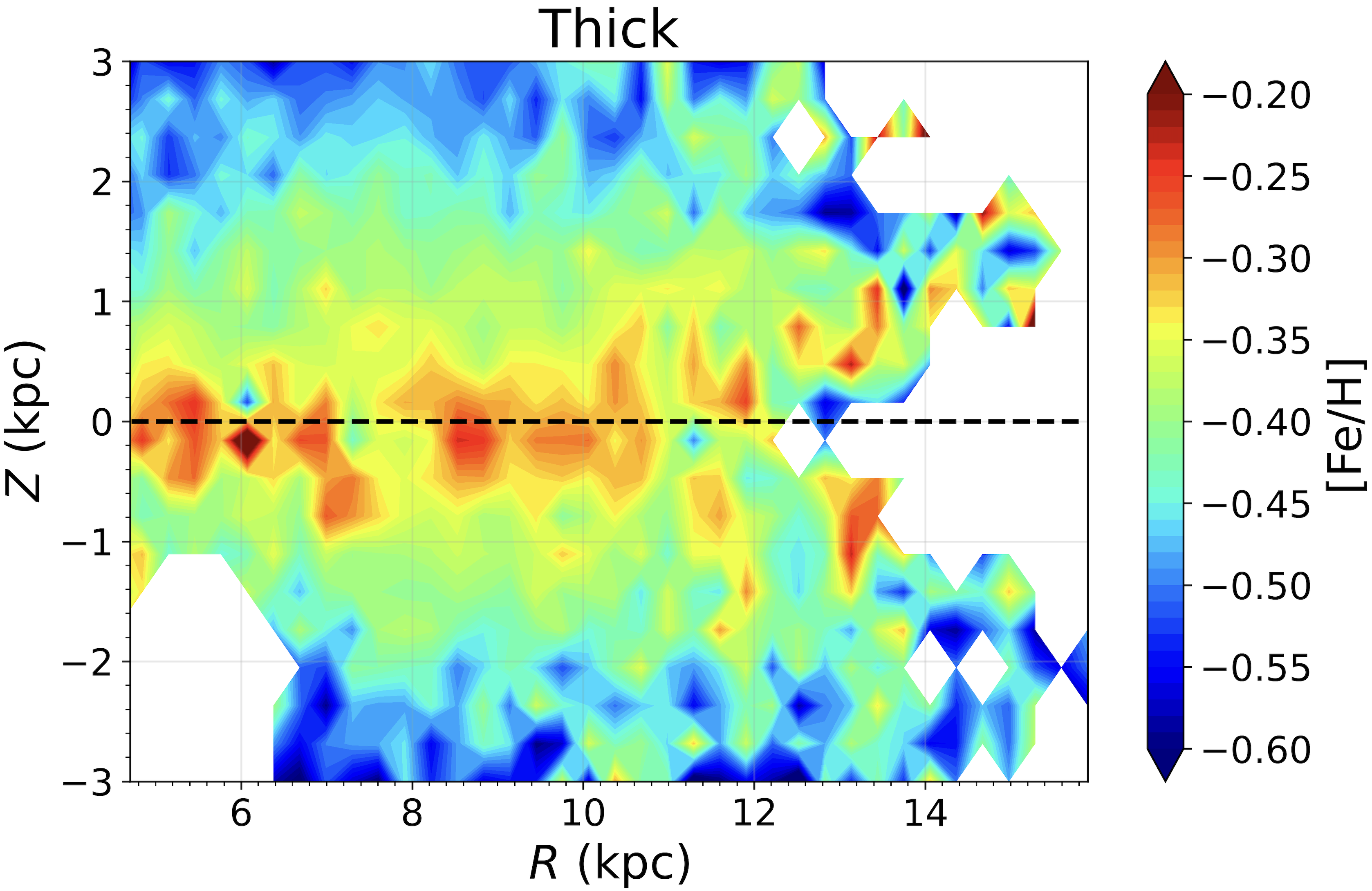}
}

\caption{Metallicity distribution, in the $R$ - $Z$ plane, of the thin (left panel) and thick (right panel) disks.}
\end{figure*}

\begin{figure*}[t]
\centering
\subfigure{
\includegraphics[width=15.cm]{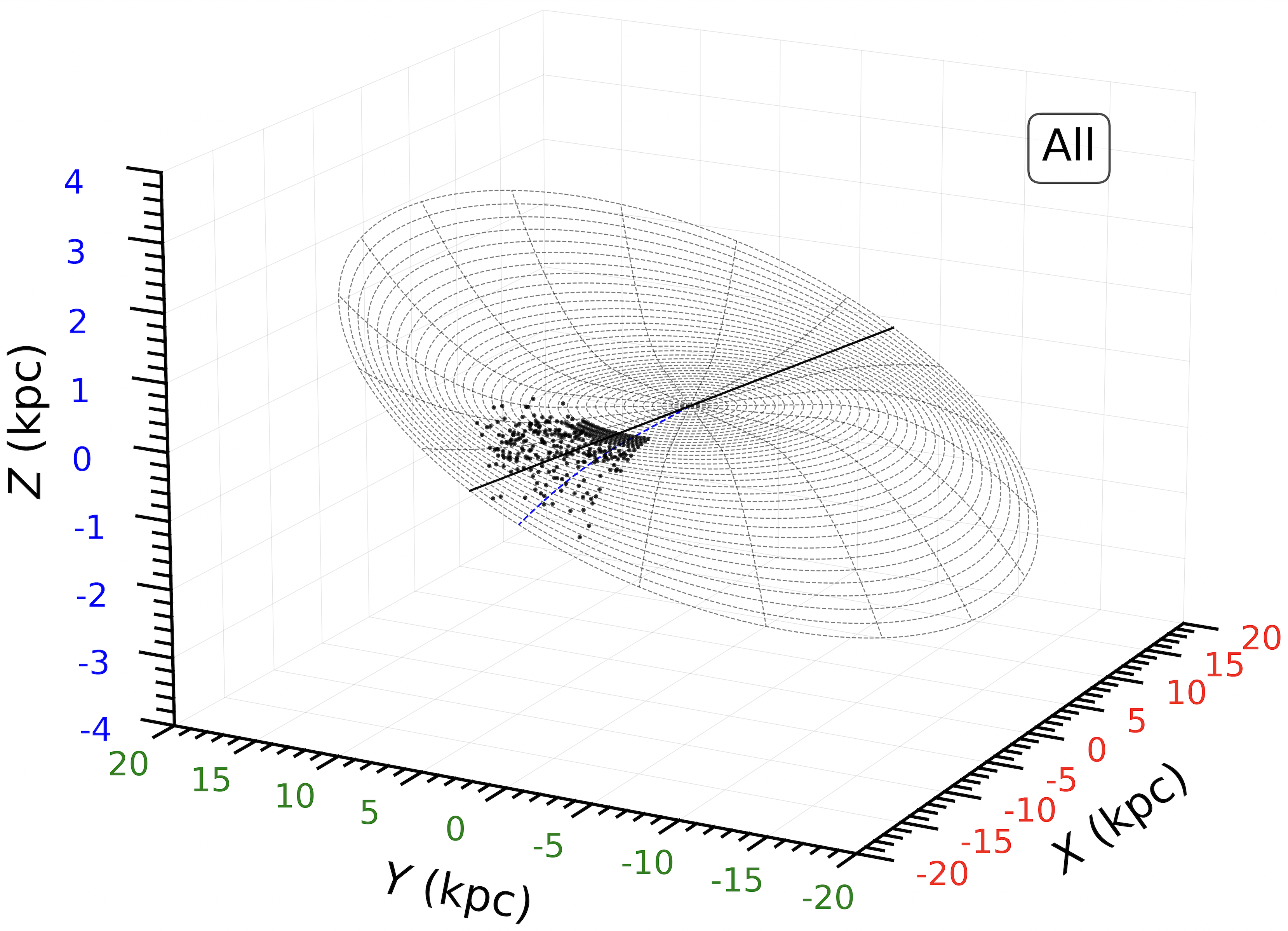}
}
\caption{The disk warp is traced by the disk metallicity mid-plane, of the whole RC sample stars.
The black dots represent the disk mid-plane determined by the method of the right panel of Fig.\,3.
The grid is the best fit with Equation (1), that is, $Z_{w}$ = 0.017\,($R$\,$-$\,7.112)$^{2}$\,sin($\phi$\,$-$\,9.218).
The black solid line indicates the LON, 9.218\,$_{-1.636}^{+1.689}$ degree, and the blue dashed line is the radius of the disk warp at Sun–Galactic Center direction ($\phi$ = 0 degree).
The LON obviously deviates from the Sun–Galactic Center direction.}
\end{figure*}

\begin{figure*}[t]
\begin{center}
\subfigure{
\includegraphics[width=15.cm]{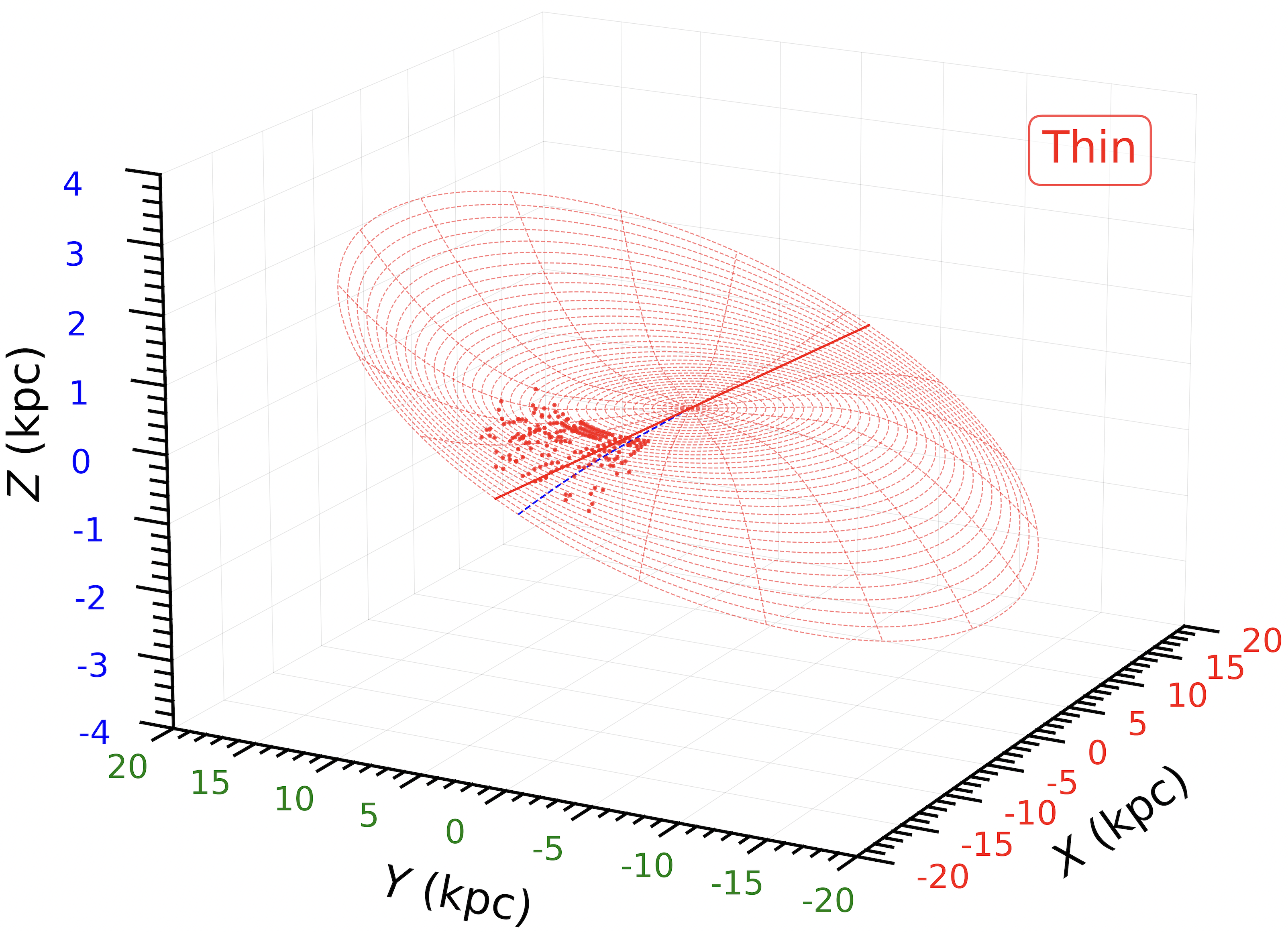}
}
\end{center}

\caption{The disk warp is traced by the disk metallicity mid-plane of the thin disk stars.
The red dots represent the disk mid-plane determined by the method of the right panel of Fig.\,3.
The grid is the best fit with Equation (1), that is, $Z_{w}$ = 0.016\,($R$\,$-$\,6.507)$^{2}$\, sin($\phi$\,$-$\,4.240).
The red solid line indicates the LON, $\phi_{0}$ = 4.240\,$_{-1.747}^{+1.641}$ degree, and the blue dashed line marks the radius of the disk warp at Sun–Galactic Center direction ($\phi$ = 0 degree).
}

\end{figure*}

In this paper, we used 171,320 red clump (RC) stars \citep{Sun2024b} from APOGEE \cite{Majewski2017} and LAMOST \citep{Deng2012, Cui2012, Liu2014, Yuan2015} surveys.
By selecting the parameters with a large signal-to-noise ratio (SNR) for the common sources, where 39,675 RC stars from APOGEE \citep{Bovy2014} and 137,448 RC stars from LAMOST \citep{Huang2020}.
The uncertainties of the effective temperature ($T_{\rm eff}$), surface gravity (log$_{g}$), line-of-sight velocity ($V_{\rm r}$),  [$\alpha$/Fe] and [Fe/H], of the whole sample stars, are typically better than, 100\,K, 0.10\,dex, 5\,km s$^{-1}$, 0.03$-$0.05\,dex and 0.10$-$0.15\,dex, respectively \citep{Bovy2014, Huang2020}.
Given the distance from the standard-candle nature of RC stars, their typical accuracies are generally better than 5\%--10\% \citep{Bovy2014, Huang2020}.

The standard Galactocentric cylindrical Coordinate ($R$, $\phi$, $Z$) has been used for this study, with three velocity components respectively $V_{R}$, $V_{\phi}$ and $V_{z}$.
$R$ is the projected Galactocentric distance with increasing radial outwards, $\phi$ is the Galactic disk rotational direction, and $\phi$ = 0$^{\circ}$ points to the Galactic anti-center direction, $Z$ points to the North Galactic Pole.
To transform the heliocentric coordinates to Galactocentric coordinates, we set the local circular velocity as $V_{c,0}$ = 238 km s$^{-1}$ \citep{Reid2004, Schonrich2010, Schonrich2012, Reid2014, Huang2015, Huang2016, Bland-Hawthorn2016}, the Galactocentric distance of the Sun as $R_{\odot}$ = 8.34 kpc \citep{Reid2014}, and solar motions as ($U_{\odot}$, $V_{\odot}$, $W_{\odot}$) $=$ $(13.00, 12.24, 7.24)$ km s$^{-1}$ \citep{Schonrich2018}.

To ensure the accuracy of the chemical distribution calculations, we further use cuts with SNRs $>$ 20 and the distance uncertainty $\leq$ 10\%.
To exclude any possibility of the halo stars, we further set stellar [Fe/H] $\geq -1.0$ dex and $\ |V_{z}|$ $\leq$ 120 km s$^{-1}$ \citep{Huang2018, Hayden2020, Sun2020}.
Finally, 170,729 RC stars were selected, of which 39,112 and 131,617 stars from the APOGEE and LAMOST surveys, respectively.
The spatial distribution of the final selected stars is shown in Fig.\,1.

Since the stellar parameters, such as [Fe/H] and [$\alpha$/Fe], are distinguished for these two surveys, considering that the LAMOST has a larger size than the APOGEE, and therefore, these parameters of APOGEE datasets are calibrated to the LAMOST datasets based on the best linear fit of the conman targets of the two surveys.

Given that the Galactic warp of the younger/thin disk is significantly stronger than that of the old/thick disk \citep[e.g.,][]{Li2020, Poggio2018, Sun2024a}, we separate our RC sample stars into thin and thick disks based on their locations on the [Fe/H]--[$\alpha$/Fe] plane in Fig.\,2.
To get clean thin/thick disk stars, two empirical cuts are used to separate the two disks \citep{Bensby2011, Lee2011, Han2020, Sun2020}.
The lower and higher-[$\alpha$/Fe] populations in this plot are respectively thin and thick disks, with 135,009 and 23,168 stars.

\section{Results and Discussion}

The results of the metallicity distribution for various populations are shown in Fig.\,3--4.
We can find that the remaining metallicity ($\Delta$[Fe/H], defined as the [Fe/H] at each position minus the average [Fe/H] at its Galactocentric radius) distribution in $R$--$Z$ plane (see the right panel of Fig.\,3).
This plot visually displays the Galactic disk metal-rich plane as a function of $R$.
The fact that the results \citep{Schonrich2017, Sun2024b} of chemo-dynamics have generally revealed that at each $R$, metal-rich stars tend to be distributed at the mid-plane of the Galactic disk.
Therefore, we approximate the plane defined by metal-rich stars as the mid-plane of the stellar disk.
The result indicates that the $Z$ height of the mid-plane (hereafter $Z_{m}$) defined as the metal-rich stars increases as $R$ increases, which displays a shape similar to the disk warp structure.
The red line in the Fig.\, 3 is the warp model determined by COBE/DIRBE data\citep{Drimmel2001}, that is, $Z_{w}$($R$)\,=\,0.0274\,($R$\,$-$\,$R_{w}$)$^{2}$sin($\phi$), here we assuming sin($\phi$) =1.0 since almost of our sample stars outer the solar radius are distributed around the Galactic anti-center direction (see Fig.\,1).
We can find that structurally, the $Z_{m}$ determined by the metal-rich stars is highly similar to the disk warp (red line) by COBE/DIRBE data \citep{Drimmel2001}, meaning that the disk south-north asymmetry could be strongly related to the disk warp \citep{Momany2006, Poggio2018, Mackereth2019, Sun2024b}.

To further support our hypothesis, we make a detailed measurement of the $Z_{m}$ as functions with $R$ for different $\phi$ bins for our RC sample stars.
The $\phi$ bins have a width of 5.0\,degree from $-$10.0\,degree to 40.0\,degree, with a 3.0\,degree overlap between adjacent $\phi$ bin.
For i-th $\phi$ bin ($\phi$ = [$\phi_{i}$, $\phi_{i}$+ 5.0]\,degree), we use the method of Fig.\,3 to measure the [Fe/H] and $\Delta$[Fe/H] distributions in the $R$--$Z$ plane, and set $R$ bins to have width of 0.5\,kpc to measure the $Z_{m}$ as a function of $R$, and set ($\phi_{i}$\,+\,$\phi_{i}$+ 5)/2.0\,degree as the unified $\phi$ angle for all data points in this $\phi$ bin.
The $Z_{m}$ at each $R$ bin represents the $Z$ at which [Fe/H] is maximum, determined by fitting the [Fe/H]--$Z$ profile of stars in this $R$ bin.
We also excluded data points that might be in areas significantly affected by the perturbation from recent Sagittarius Passing by \citep[e.g.,][]{Sun2024a, Sun2024c}.
The determined data points are marked as black dots in Fig.\,5, the result indicates that all data points form a warped plane, and hence, we fitted the results using a common warp model \citep{Drimmel2001, Poggio2017, Chen2019} as follows:
 \begin{equation}
     Z_{w} =
     \begin{cases}
     A_{w}\,(R - R_{w})^{2}\,\mathrm{sin}(\phi - \phi_{0}), & R > R_{w},\\
     0, & R \le R_{w}.
     \end{cases}
 \end{equation}
Where the $Z_{w}$, $A_{w}$, $R_{w}$ and $\phi_{0}$, are respectively, the $Z$ height, amplitude, onset radius and LON of the warp.

We use the Python package emcee\cite{Foreman-Mackey2013} to produce an MCMC sampling of the posterior distribution of the two parameters, defining the minimum of the least squares as the negative log-likelihood and setting 100 “walkers,” running 5000 steps by removing the first 2000 steps for each walker.
The corner plots of the posterior distribution of the MCMC samples of the two parameters (and their 1$\sigma$ confidence intervals) are displayed in Appendix Fig.\,A1.

The best fit is displayed by the black plots in Fig.\,5, the results of which yields $Z_{w}$ = 0.017\,($R$\,$-$\,7.112)$^{2}$\,sin($\phi$\,$-$\,9.218).
Our result indicates that the LON is not oriented in the Galactic Center-Sun direction (the blue dashed line in the figure is the radius of the disk warp at Galactic Center-Sun direction), instead, it points an angle around 9.218\,$_{-1.636}^{+1.689}$ degree, and the onset radius is around 7.112$^{+0.654}_{-0.641}$\,kpc.
This result is in good agreement with the intuitive measurement of the disk warp based on the Cepheids samples\citep[e.g.,][]{Chen2019, Huang2024}, which suggested the $\phi_{0}$ = 10.06 $\pm$ 0.93 degree \citep{Huang2024} and $R_{w}$ = 7.70$^{+0.34}_{-0.42}$\,kpc \citep{Chen2019}.
The parametrization presented by our results exhibits a maximum warp amplitude of about $\sim$0.8\,kpc at $R$ $\sim$ 13--14\,kpc, which is slightly smaller than that of other studies based on RC samples \citep[e.g.,][]{Uppal2024, Khanna2024}, while it is in perfect agreement with the warp amplitude determined by Cepheids/young giants \citep[e.g.,][]{Chen2019, Dehnen2023, Cabrera-Gadea2024, Poggio2024}.
The morphology consistency implies the disk warp is likely the origin of the disk north-south asymmetry in [Fe/H].
Therefore, the disk north-south asymmetry in [Fe/H] should be a new tracer for the disk warp, which will overcome many shortcomings brought by the use of other tracers \citep[e.g.,][]{Huang2018, Chen2019, Sun2024a}, including:
(i) It avoids the problem of the stellar kinematic tracers requiring one to assume a specific kinematic model to connect to the structural properties of the warp;
(ii) It overcomes the disadvantage of correcting the selection effect of intuitive stellar density tracers.

Given that the south-north asymmetry in [Fe/H] of the thin disk is significantly stronger than that of the thick disk (see Fig.\,4), to avoid any possible effect from the larger contamination of thick disk stars in the RC sample and thereby obtaining a more accurate measurement of the Galactic warp morphology, we further use the disk metallicity mid-plane to trace the warp of the thin disk, and the result is shown in Fig.\,6.
We can easily find that the thin disk result also forms a warped shape (Fig.\,6), and hence, we also fit the shape of the disk warp of the thin disk stars with Equation (1).

The best fit is displayed by the red plots in Fig.\,6, the results yield the thin disk warp is accurately described as $Z_{w}$ = 0.016\,($R$\,$-$\,6.507)$^{2}$\, sin($\phi$\,$-$\,4.240).
The LON with 4.240\,$_{-1.747}^{+1.641}$ degree is slightly larger than the result of COBE/DIRBE data\citep{Drimmel2001}, and is still not oriented in the Galactic Center-Sun direction, while it is in rough agreement with the young ($\sim$20--120\,Myr) Cepheid sample with $\phi_{0}$ = 6.14$\pm$1.34 degree as reported by Huang et al. ({\color{blue}{2024}}).
It is worth noting that the measurements of the LON presented by Huang et al. ({\color{blue}{2024}}) are prone to an omitted-variable bias, which is mainly because the classical Cepheids exhibit a positive correlation between age and distance from the Galactic center \citep[e.g.,][]{Skowron2019, Anders2024}.
This feature is specific to Cepheids and is due to a metallicity dependence of the instability strip.
Because of this property, binning Cepheids in age also implies binning in Galactocentric radius (i.e., considering only a part of the disk).﻿
While this would not pose an issue if the LON of the warp were straight, it has been shown that the LON gradually varies as a function of Galactocentric radius \citep[e.g.,][]{Dehnen2023, Cabrera-Gadea2024, Poggio2024}.
Thus, comparing our results to those derived from the young Cepheids in Huang et al.({\color{blue}{2024}}), it is effectively equivalent to comparing the LON in a given portion of the Galactic disk.
Here, we would like to clarify that the simple model used in this study assumes a straight LON.
Given this assumption, when comparing our results to other works in the literature based on Cepheids, it may be necessary to consider the impact of the radial dependence of the LON of Cepheids.
Therefore, it is useful to compare our results to a range of LON values determined by the Cepheids (instead of just a single number) in the approximate radial range covered by our dataset.
We can find that our determined LON value falls well within the range of LON values for the Cepheids in the approximate radial range covered by our dataset \citep[e.g.,][]{Dehnen2023, Poggio2024}.

It is worth emphasizing that inferring the shape of the Galactic warp is a notoriously difficult task.
This is mainly due to: (i) Extinction, which prevents us from looking at some regions of the sky, especially in the Galactic plane \citep[e.g.,][]{Chen2019, Green2019, Lallement2022};
(ii) Selection function, the catalog of the LAMOST/APOGEE dataset is not complete, the sampling of the which has a very inhomogeneous selection function, only specific line-of-sights are considered, and others are completely absent \citep[e.g.,][]{Majewski2017, Mackereth2019, Huang2020}.
These limitations are severely limiting the quality of our inferred warp shape.
We encourage further observational efforts to obtain more comprehensive data, which will facilitate more accurate inferences regarding the 3D shape of the Galactic warp.

\section{Conclusions}
Using a sample with over 170,000 RC stars selected from LAMOST and APOGEE, we measure the [Fe/H] distribution for various populations, across a larger disk volume (4.0\,$\leq$\,$R$\,$\leq$15.0\,kpc and $|Z|$\,$\leq$3.0\,kpc).
We find that:
\\
\\
$\bullet$ The Galactic disk displays obvious north-south asymmetry in [Fe/H], and the height of the mid-plane of the [Fe/H] disk for the whole RC sample stars is accurately described as $Z_{w}$ = 0.017\,($R$\,$-$\,7.112)$^{2}$\,sin($\phi$\,$-$\,9.218), with the line of node (LON) is oriented at 9.218\,$_{-1.636}^{+1.689}$ degree.
This result in morphology is in good agreement with the stellar warp traced by Cepheids, meaning that the disk north-south asymmetry in [Fe/H] may be related to the disk warp.
\\
\\
$\bullet$ The result of the young/thin disk stars of this RC sample indicates that its warp is well-modeled as $Z_{w}$ = 0.016\,($R$\,$-$\,6.507)$^{2}$\, sin($\phi$\,$-$\,4.240).
This result indicates that the LON of the young/thin disk stars is around $\phi_{0}$ = 4.240\,$_{-1.747}^{+1.641}$ degree, which is again in rough agreement with the young ($\sim$20--120\,Myr) Cepheid sample result.
\\
\\
The consistency in morphology indicates that the disk north-south asymmetry in [Fe/H] may be related to the disk warp, and should be a new tracer for the disk warp.
The accurate Galactic warp measurement by thin disk stars indicates that its LON is $\phi_{0}$ = 4.240\,$_{-1.747}^{+1.641}$ degree.

\section*{Acknowledgements}
We thank the anonymous referee for very useful suggestions to improve the work. This work is supported by the NSFC projects 12133002, 11833006, and 11811530289, and the National Key R\&D Program of China No. 2019YFA0405500, 2019YFA0405503, and CMS-CSST-2021-A09, and the Postdoctoral Fellowship Program of CPSF under Grant Number GZC20240125, and the China Postdoctoral Science Foundation under Grant Number 2024M760240.

It is a pleasure to thank Zhaoyu Li (SJTU), Maosheng Xiang (NAOC), Yang Huang (UCAS), and Hao Tian (NAOC) for their helpful discussions.

Guoshoujing Telescope (the Large Sky Area Multi-Object Fiber Spectroscopic Telescope LAMOST) is a National Major Scientific Project built by the Chinese Academy of Sciences. Funding for the project has been provided by the National Development and Reform Commission. LAMOST is operated and managed by the National Astronomical Observatories, Chinese Academy of Sciences.

\bibliographystyle{aasjournal}

\appendix

\section{Example posterior distributions of MCMC samples of the parameters in warp model of different populations}

This appendix present the example posterior distributions of the parameters in warp model with Equation (1), of the whole sample stars (left panel of Fig.\,{\color{blue}{A1}}) and the thin disk stars (right panel of Fig.\,{\color{blue}{A1}}).

\begin{figure*}[t]
\centering

\subfigure{
\includegraphics[width=8.5cm]{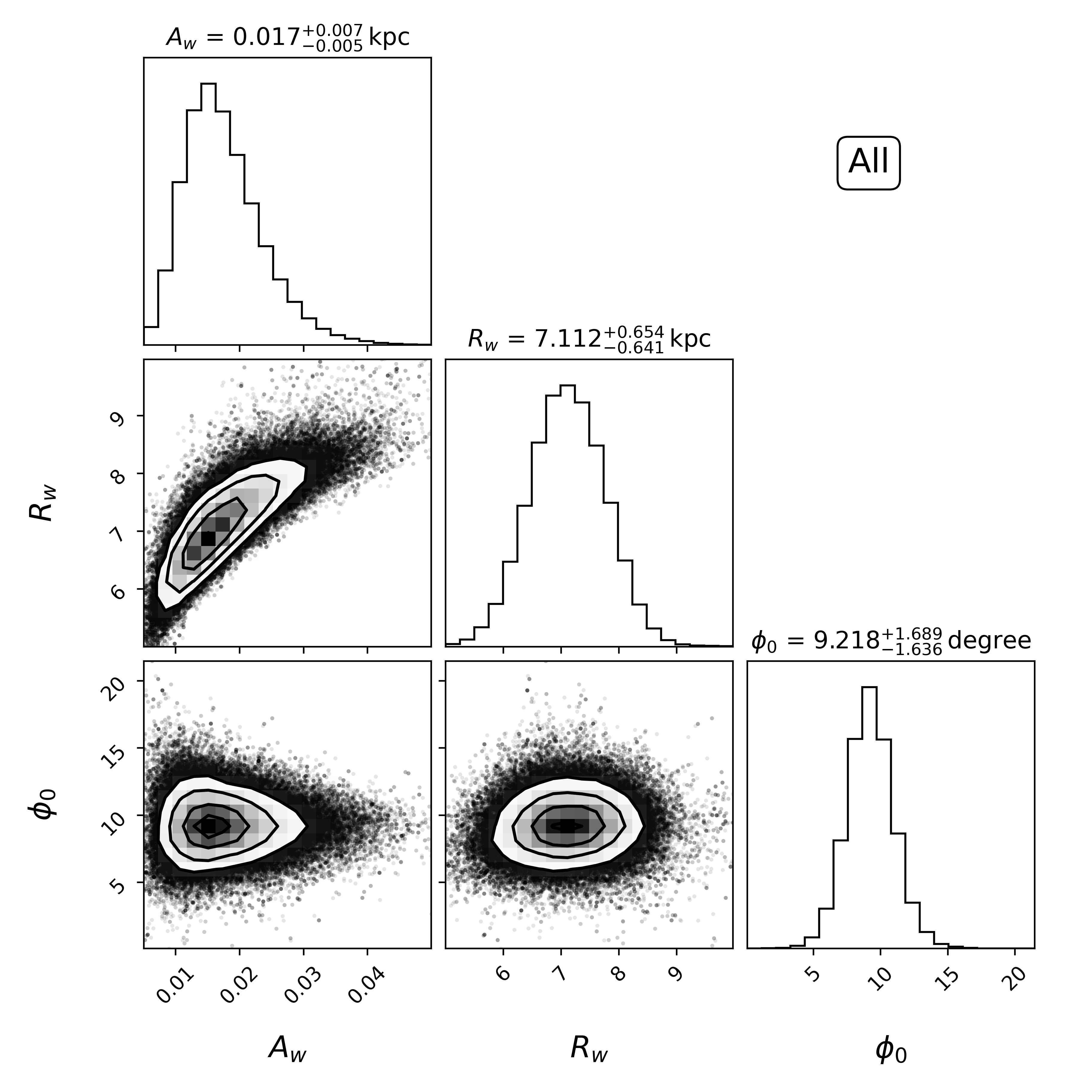}
}
\subfigure{
\includegraphics[width=8.5cm]{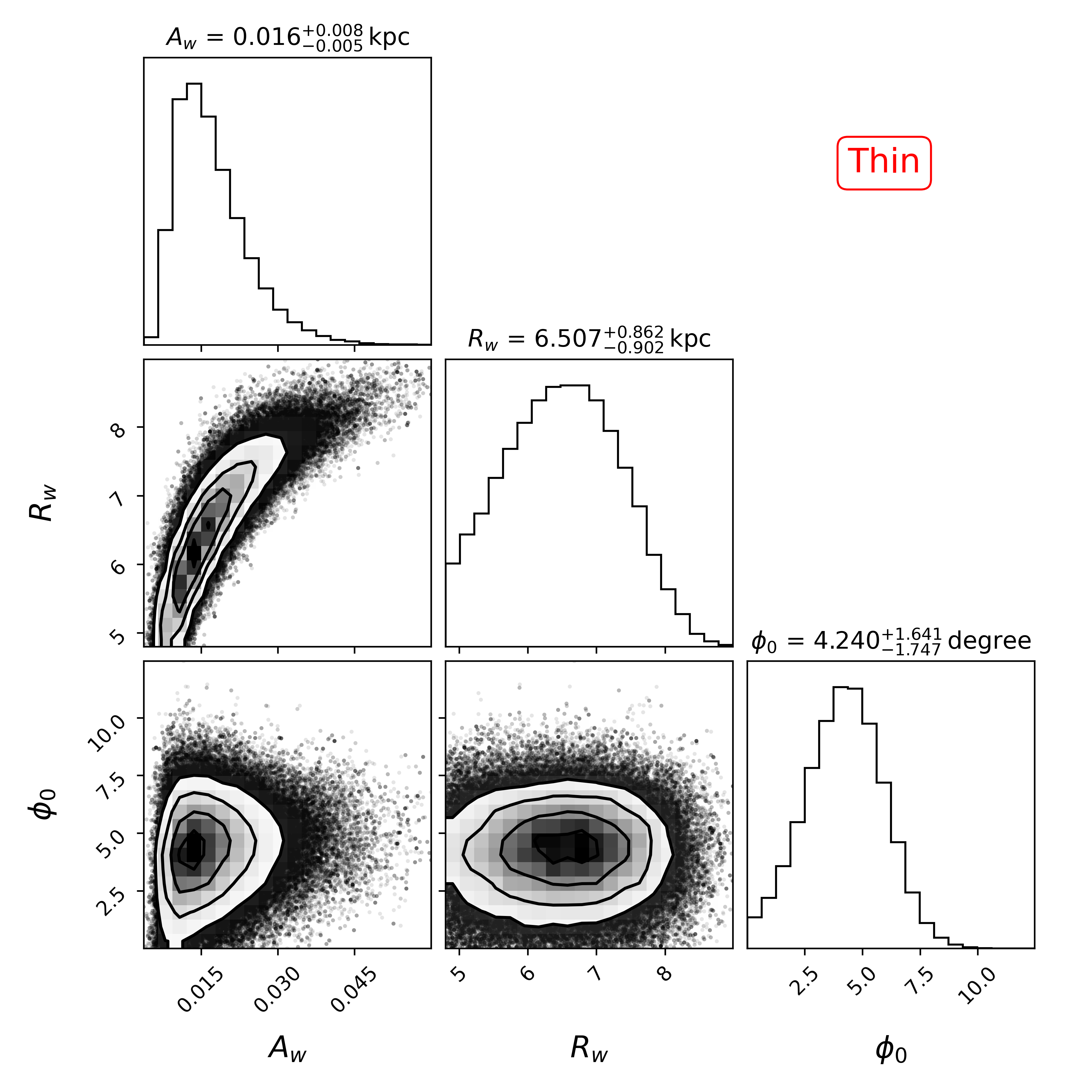}
}

\caption{The corner plots of posterior distributions of MCMC samples for the parameters in warp model with $Z_{w}$ = $A_{w}$\,($R$ $-$ $R_{w}$)$^{2}$\,sin($\phi$ $-$ $\phi_{0}$), of the whole sample stars (left panel) and the thin disk stars (right panel).
The shadowed areas circled from inside to outside represent the confidence intervals of 1$\sigma$, 2$\sigma$ and 3$\sigma$, respectively.}
\end{figure*}


\end{document}